\documentclass[trackchanges]{aastex701}

\usepackage{hyperref}
\usepackage{graphicx}
\usepackage{dcolumn}
\usepackage{bm}
\usepackage{epsfig}
\usepackage{amsmath}
\usepackage{ulem}

\usepackage[T1]{fontenc} 
\usepackage[utf8]{inputenc} 
\usepackage[spanish, english]{babel} 

\begin{document}

\title{Cosmological evolution of fast radio bursts and its rapid decline relative to star formation rate}

\author[0009-0009-3583-552X]{X. D. Jia}
\affiliation{School of Astronomy and Space Science, Nanjing University, Nanjing 210093, China}
\email{602023260009@smail.nju.edu.cn}

\author[0000-0001-7176-8170]{D. H. Gao}
\affiliation{School of Astronomy and Space Science, Nanjing University, Nanjing 210093, China}
\email{2300761756@qq.com}

\author[]{J. H. Chen}
\affiliation{School of Astronomy and Space Science, Nanjing University, Nanjing 210093, China}
\email{2630578993@qq.com}

\author[0000-0001-6021-5933]{Q. Wu}
\affiliation{School of Astronomy and Space Science, Nanjing University, Nanjing 210093, China}
\email{wqin@smail.nju.edu.cn}

\author[0000-0003-0672-5646]{S. X. Yi}
\affiliation{School of Physics and Physical Engineering, Qufu Normal University, Qufu 273165, China}
\email{yisx2015@qfnu.edu.cn}

\author[0000-0003-4157-7714]{F. Y. Wang}
\affiliation{School of Astronomy and Space Science, Nanjing University, Nanjing 210093, China}
\affiliation{Key Laboratory of Modern Astronomy and Astrophysics (Nanjing University), Ministry of Education, China}
\email{fayinwang@nju.edu.cn}

\correspondingauthor{F. Y. Wang}
\email{fayinwang@nju.edu.cn}

\begin{abstract}
Fast radio bursts (FRBs) are enigmatic millisecond-duration radio transients whose physical origins remain debated. To shed light on this, we analyze the CHIME/FRB Catalog 2. By using the probability distribution of dispersion measured (DM) derived from the IllustrisTGN simulation, we compute the pseudo-redshift with $1\sigma$ error for each FRB. To derive the FRB luminosity function and event rate, we employ a non-parametric statistical method. Building upon Efron-Petrosian method, we find strong luminosity evolution with redshift, well described by $L_0 \propto (1+z)^{6.38}$. After de-evolving this trend, we apply Lynden-Bell's $C^-$ method to derive the comoving FRB formation rate which is found to decline rapidly at high redshift, following $\rho(z) \propto (1+z)^{-5.38 \pm 0.02}$. We also test the robustness of our results bu considering the upper and lower limits of pseudo-redshifts, and different flux limits of CHIME. Similar results are found. This steep decline is inconsistent with a direct tracing of the cosmic star formation rate, but closely resembles the redshift evolution of short gamma-ray bursts—systems linked to compact object mergers. Our results support that the origin of FRBs is associated with old populations, such as neutron stars and black holes.
\end{abstract}



\section{Introduction} 
Fast radio bursts (FRBs) are an extremely bright class of millisecond-duration radio transients originating in the universe, which were first discovered in 2007 \citep{2007Sci...318..777L}. Subsequently, multiple similar radio pulses were detected one after another, drawing widespread attention in the scientific community to FRBs as a high-energy astrophysical phenomenon \citep{2012MNRAS.425L..71K,2013Sci...341...53T}. To date, with the improvement of observational techniques, thousands of FRB events have been discovered \citep{2021ApJS..257...59C,2026arXiv260109399F}. Although their physical origin remains uncertain, FRBs are generally thought to originate outside the Milky Way (MW) due to their high burst rate and large dispersion measure (DM) \citep{2021SCPMA..6449501X,2022A&ARv..30....2P,2023RvMP...95c5005Z,2024ChPhL..41k9801W}. Subsequent identification of host galaxies and redshift measurements have further confirmed this expectation \citep{2017Natur.541...58C,2017ApJ...834L...7T}, with only one FRB to date conclusively traced to a source within our Galaxy \citep{2020Natur.587...54C}. 


The physical origin of FRBs remains uncertain. Several theoretical models have been proposed to explain FRBs \citep{Platts2019}. The leading source model is magnetar. 
The discovery of FRB 20200428 associated with the magnetar SGR 1935+2154 supports that at least a subset of FRBs originate from magnetars formed in the deaths of massive stars \citep{2020Natur.587...59B,2020Natur.587...54C}. Recent studies found that young magnetars are required to explain the observations of some active FRBs \citep{Wang2022,Wang2025}. Therefore, the view that FRBs trace the cosmic star formation history has been widely adopted. However, a repeating source FRB 20200120E was found to be associated with the nearby spiral galaxy M81 at a distance of $3.6$ Mpc \citep{2021ApJ...910L..18B}. Follow-up observations surprisingly revealed that the source is located in a globular cluster within the host galaxy \citep{2022Natur.602..585K}. It suggests that some FRBs are linked to ancient stellar populations rather than being directly associated with young populations. 

In addition to directly observing FRB sources, statistical analysis is used to infer their progenitors from the event rate. By analyzing samples from Parkes and Australian Square Kilometer Array Pathfinder (ASKAP), \cite{2019MNRAS.487.3672Z} studied the energy function and formation rate of FRBs, taking into account the effects of metallicity. They found that FRB formation exhibits a moderate time delay -- approximately $3\sim5$ Gyr -- relative to the cosmic SFR. \cite{2024A&A...690A.377W} also reached a similar conclusion: the FRB event rate evolves with star formation rate (SFR), but with a short delay time \citep{Gupta2025}. Moreover, constraints on the origin of FRBs can also be placed by comparing the spatial number density of FRB sources with that of potential progenitor populations \citep{2018ApJ...858...89C,2019NatAs...3..928R,2020MNRAS.494.2886H,2020MNRAS.494..665L}. Studies of the luminosity function or energy distribution of FRBs have yielded constraints on their progenitor systems \citep{2020MNRAS.494.2886H,2021MNRAS.501.5319A,2022MNRAS.510L..18J,2022MNRAS.511.1961H,Wu2025ApJ}. Based on the first CHIME/FRB catalog, \cite{2022ApJ...924L..14Z} and \cite{Qiang2022} ruled out the hypothesis that all FRBs strictly follow the SFR \citep{Lin2024}. In addition to the forward fitting method described above, non-parametric methods are also employed. Using the Lynden-Bell's $c^{-}$ method, \cite{2024ApJ...973L..54C} found strong luminosity evolution and a rapidly declining formation rate proportional to $(1+z)^{-4.9\pm0.3}$. Upon comparing this with SFR, they concluded that the origin of FRBs is associated with old populations including neutron stars and black holes. \cite{2025ApJ...988L..64C} considered different redshift values and forms of redshift evolution, and consistently found a time delay between the FRB formation rate and SFR. By studying the host galaxies of astrophysical sources, a possible connection between the progenitors of FRBs and type Ia supernovae is found \citep{Horowicz2026}. 

In this paper, we calculate the pseudo-redshifts of the FRBs in CHIME/FRB catalog 2 and use the Lynden-Bell's $c^{-}$ and Efron-Petrosian methods to infer their luminosity function and formation rate. In Section \ref{pseudo-redshift distribution}, we describe the FRB sample and the method used to estimated pseudo-redshifts. Section \ref{Lynden-Bell's $C^{-}$ method} presents the Lynden-Bell's $c^{-}$ and Efron-Petrosian methods. In Section \ref{Luminosity function and Formation rate}, we show the resulting luminosity function and formation rate of FRBs. Finally, Section \ref{Conclusions and discussion} summarizes our conclusions and discussion.

\section{redshift distribution}\label{pseudo-redshift distribution}
\subsection{FRB Sample}\label{FRB Sample} 
CHIME has discovered hundreds of FRBs since 2018 \citep{2021ApJS..257...59C}. Recently, CHIME released the largest FRB catalog to date, containing a total of 4,539 FRB events detected between July 2018 and September 2023, referred to as "Catalog 2" \citep{2026arXiv260109399F}. This new Catalog 2 includes the events from old Catalog 1 and has been uniformly reprocessed using an improved analysis pipeline \citep{2021ApJS..257...59C,2026arXiv260109399F}. Based on Catalog 2, we can infer the pseudo-redshifts of FRBs and calculate their luminosity function and formation rate.

The Catalog 2 comprises 4,539 bursts, including apparently 3,558 one-off bursts and 981 repeating bursts from 83 sources 2. In this paper, we primarily focus on the 3,558 one-off bursts and apply a selection criterion to filter them. Among these data, all bursts have well measured DM, we perform a selection based on the extragalactic contribution to DM ($\mathrm{DM}_{\mathrm{exc}}$) and the Galactic DM component ($\mathrm{DM}_{\mathrm{MW}}$). First, since the DM contribution of Galactic halo is approximately $50 \sim 80$ pc cm$^{-3}$, we remove FRBs with $\mathrm{DM}_{\mathrm{exc}} < 100$ pc cm$^{-3}$ \citep{2023ChPhC..47h5105T}. Second, to mitigate the large uncertainties of the Galactic DM contribution near the Galactic plane, we remove FRBs with $\mathrm{DM}_{\mathrm{MW}} > 200$ pc cm$^{-3}$ \citep{2025arXiv250608932W}. Finally, we remove FRBs that simultaneously satisfy $\mathrm{DM}_{\mathrm{MW}} > 100$ pc cm$^{-3}$ and $\mathrm{DM}_{\mathrm{MW}} > \mathrm{DM}_{\mathrm{exc}}$. The final sample contains 3,293 FRBs. In addition to DM, we also apply selection criteria based on their flux. First, we remove FRBs without flux measurements. Second, we exclude bursts whose flux uncertainties are larger than their measured flux values. This leaves us with a final sample of 2,982 FRBs. Here, we adopt two flux limits as $0.2$ Jy and $0.5$ Jy \citep{2024ApJ...973L..54C,2025ApJ...988L..64C}. Among these FRBs, the sample with flux larger than $0.2$ Jy contains 2,941 bursts, and the sample with flux larger than $0.5$ Jy contains 2,230 bursts.

\subsection{The contribution of DM}\label{The contribution of DM}

Due to the propagation speed of electromagnetic waves in a plasma depends on frequency, a signal that is emitted instantaneously at the source arrives at Earth with its lower-frequency components delayed relative to its higher-frequency components. This phenomenon is known as dispersion. The observed DM is an important observational quantity for FRBs. It is typically divided into the following four components: Milky Way interstellar medium (DM$_{\textrm{MW}}$), galactic halo (DM$_{\textrm{halo}}$), intergalactic medium (DM$_{\textrm{IGM}}$), and host galaxy (DM$_{\textrm{host}}$),
\begin{equation}
    \mathrm{DM}_{\mathrm{obs}} = \mathrm{DM}_{\mathrm{MW}} + \mathrm{DM}_{\mathrm{halo}} + \mathrm{DM}_{\mathrm{IGM}} + \frac{\mathrm{DM}_{\mathrm{host}}}{1+z}.\label{eq:DM}
\end{equation}
Here, $\mathrm{DM}_{\mathrm{obs}}$ is the total observed DM, $\mathrm{DM}_{\mathrm{host}}$ is the DM of the host galaxy in the FRB's source frame, with the factor $(1+z)$ accounts for time dilation. $\mathrm{DM}_{\mathrm{MW}}$, the DM contributed by the interstellar medium of the MW, is well described by Galactic electron distribution models such as YMW16 and NE2001 \citep{2002astro.ph..7156C,2017ApJ...835...29Y}. The component $\mathrm{DM}_{\mathrm{halo}}$ is currently poorly constrained. In this work, we adopt a conservative estimate, assuming that $\mathrm{DM}_{\mathrm{halo}}$ follows a Gaussian distribution with $\langle \mathrm{DM}_{\mathrm{halo}} \rangle$ =
$65~\mathrm{pc~cm^{-3}}$ and $\sigma = 15~\mathrm{pc~cm^{-3}}$ \citep{2019MNRAS.485..648P,2020Natur.581..391M,Wu2022}. Therefore, for simplicity, the first two terms of the right-hand side of Equation (\ref{eq:DM}) are typically subtracted to obtain the extragalactic component, expressed as:
\begin{equation}
    \mathrm{DM}_{\mathrm{exc}} \equiv \mathrm{DM}_{\mathrm{obs}} - \mathrm{DM}_{\mathrm{MW}} - \mathrm{DM}_{\mathrm{halo}} = \mathrm{DM}_{\mathrm{IGM}} + \frac{\mathrm{DM}_{\mathrm{host}}}{1+z}.
\end{equation}

Within the framework of the $\Lambda$CDM cosmological model, the mean value of $\mathrm{DM}_{\mathrm{IGM}}$ can be expressed as 
\begin{equation}
    \langle \mathrm{DM}_{\mathrm{IGM}} \rangle = \frac{3cH_0\Omega_b f_{\mathrm{IGM}}}{8\pi G m_p} \times f_e(z),
\end{equation}
where $c$ is the speed of light, $H_0$ is the Hubble constant, $\Omega_b$ is the densities of baryon matter, $f_{\mathrm{IGM}}$ is the fraction of baryons in IGM, G is the Newtonian gravitational constant, $m_p$ is the proton mass, and $f_e(z)$ is defined as 
\begin{equation}
    f_e(z) = \int_0^z \frac{\left[ \frac{3}{4} y_1 \chi_{e,\mathrm{H}}(z) + \frac{1}{8} y_2 \chi_{e,\mathrm{He}}(z) \right] (1+z)\,dz}{\left[ \Omega_m (1+z)^3 + \Omega_\Lambda \right]^{1/2}}.
\end{equation}
The parameters $y_1$ and $y_2$ are the hydrogen and helium fractions normalized to $0.76$ and $0.24$, respectively, which can be neglected as $y_1 \simeq y_2 \simeq 1$. In the late universe ($z<3$), hydrogen and helium can be considered fully ionized \citep{2009RvMP...81.1405M,2011MNRAS.410.1096B}. Therefore, their ionization fractions $\chi_{e,\mathrm{H}}(z)$ and $\chi_{e,\mathrm{He}}(z)$ can be taken as $\chi_{e,\mathrm{H}}(z) = \chi_{e,\mathrm{He}}(z) = 1$. Finally, the expression for $\mathrm{DM}_{\mathrm{IGM}}$ can be simplified to
\begin{equation}
    \langle \mathrm{DM}_{\mathrm{IGM}} \rangle = \frac{21 c \Omega_b H_0^2}{64 \pi H_0 G m_p} \times \int_0^z \frac{f_{\mathrm{IGM}} (1+z)\,dz}{\left[ \Omega_m (1+z)^3 + 1 - \Omega_m \right]^{1/2}}.\label{eq:DM_IGM}
\end{equation}
The value of $f_{\mathrm{IGM}} = 0.84$ is adopted \citep{2012ApJ...759...23S,Yang2022}. 
The cosmological parameters are adopted from the results of \cite{2020A&A...641A...6P}. 

It is worth noting that Equation (\ref{eq:DM_IGM}) represents the mean contribution from the IGM. For any individual FRB event, the true value will vary around the mean. Previous studies have shown that its probability distribution can be fitted with a Gaussian-like distribution as \citep{2020Natur.581..391M,2021ApJ...906...49Z}
\begin{equation}
    p_{\mathrm{IGM}}(\Delta) = A \Delta^{-\beta} \exp\left[ -\frac{(\Delta^{-\alpha} - C_0)^2}{2\alpha^2 \sigma_{\mathrm{IGM}}^2} \right], \quad \Delta = \frac{\mathrm{DM}_{\mathrm{IGM}}}{\langle \mathrm{DM}_{\mathrm{IGM}} \rangle},
\end{equation}
where $\alpha = \beta = 3$ \citep{2020Natur.581..391M}. The parameters $A$ and $C_0$ are normalization parameters chosen such that the distribution $p_{IGM}$ is properly normalized ($\int p_{\mathrm{IGM}} = 1$) and $\langle \Delta \rangle = 1$.

Due to the lack of detailed observations of FRB host galaxies, the uncertainty in $\mathrm{DM}_{\mathrm{host}}$ is relatively large. The numerical simulations show that the probability distribution of $\mathrm{DM}_{\mathrm{host}}$ follows a log-normal form \citep{2020Natur.581..391M,2020ApJ...900..170Z},
\begin{equation}
    p_{\mathrm{host}}(\mathrm{DM}_{\mathrm{host}}) = \frac{1}{\sqrt{2\pi}\, \mathrm{DM}_{\mathrm{host}}\, \sigma_{\mathrm{host}}} \exp\left[ -\frac{(\ln \mathrm{DM}_{\mathrm{host}} - \mu)^2}{2\sigma_{\mathrm{host}}^2} \right],
\end{equation}
where $\mu$ and $\sigma_{\mathrm{host}}$ are the mean and standard deviation of ln$\mathrm{DM}_{\mathrm{host}}$, respectively. Using the IllustrisTNG simulation \citep{2018MNRAS.473.4077P}, \cite{2020ApJ...900..170Z} and \cite{2021ApJ...906...49Z} determined the best-fit distribution parameters for both the $\mathrm{DM}_{\mathrm{IGM}}$ and $\mathrm{DM}_{\mathrm{host}}$. Here we adopt the best-fit values of the distribution parameters provided in their results ($e^{\mu},\sigma_{\mathrm{host}},\sigma_{\mathrm{IGM}},A$ and $C_0$), and use a monotone cubic spline interpolation to obtain the values at arbitrary redshifts.

\subsection{Calculating Pseudo-Redshifts}\label{Calculating Redshift}
Although a large number of FRBs have been observed to date, the majority of them remain non-localized. To better extract the cosmological information encoded in these non-localized FRBs, the observed DM can be used to infer their redshifts in reverse \citep{2025A&A...698A.215G}. To this end, we utilize a large sample of non-localized FRBs from the latest Catalog 2 to statistically infer their redshift distributions. The specific sample selection methods is described in Section \ref{FRB Sample}. Next, we calculate the pseudo-redshift for these 3,293 FRBs.

\begin{figure}
    \centering
    \includegraphics[width=0.7\textwidth]{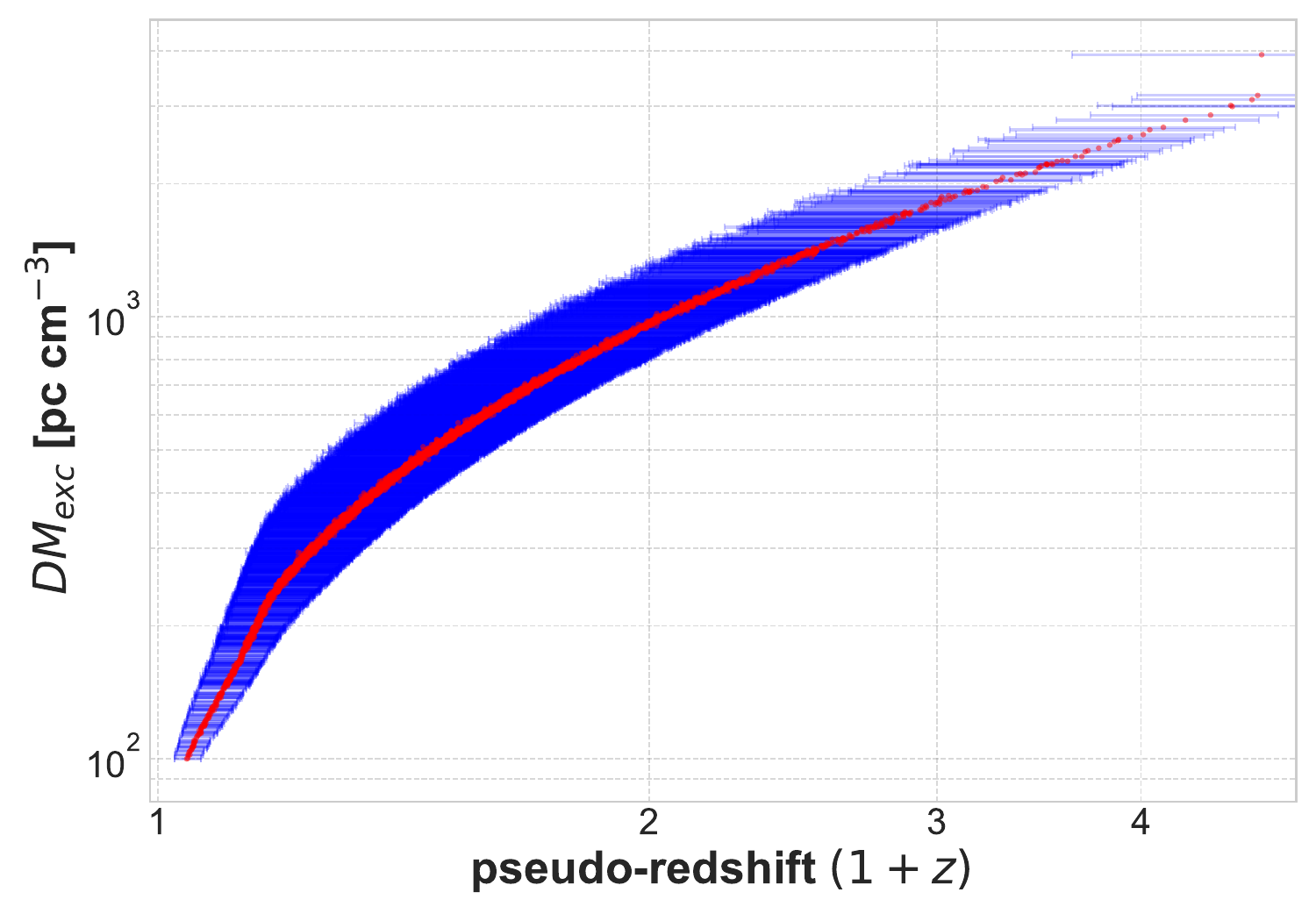} 
    \caption{Pseudo redshift distributions of non-localized FRBs. The red points with blue error bars are the best-fit values and $1\sigma$ uncertainties of redshifts estimated for each FRB.}
    \label{Fig:redshift_scatter}
\end{figure}

When calculating the redshift, all other parameters are held fixed, and only the redshift $z$ is treated as a free parameter to be fitted. Specifically, the cosmological parameters are adopted from \cite{2020A&A...641A...6P}, and the DM distributions are drawn from the probability distributions of parameters derived from the IllustrisTNG simulations \citep{2020ApJ...900..170Z,2021ApJ...906...49Z}. The likelihood function for the $i$th FRB can then be expressed as follows \citep{2025A&A...698A.215G}
\begin{equation}
\begin{split}
\mathcal{L}(z_i) = &\int_0^{(1+z_i)(\mathrm{DM}_i - \mathrm{DM}_{\mathrm{MW}})} p_{\mathrm{host}}(\mathrm{DM}_{\mathrm{host}} \mid z_i) \\
&\times\, p_{\mathrm{IGM}}\left( \mathrm{DM}_i - \mathrm{DM}_{\mathrm{MW}} - \frac{\mathrm{DM}_{\mathrm{host}}}{1+z_i} \mid z_i \right) d\,\mathrm{DM}_{\mathrm{host}}.
\end{split}
\end{equation}
We employ the Markov Chain Monte Carlo (MCMC) code $emcee$ to compute the posterior distribution of the pseido-redshift for each FRB\citep{2013PASP..125..306F}. We also calculate the pseudo-redshift based on its probability distribution, more details can be found in \cite{2025A&A...698A.215G}. The results are shown in Figure \ref{Fig:redshift_scatter}, where the blue points represent the best-fit redshift values for each FRB, with error bars indicating the $1\sigma$ uncertainties, and the red points denote the pseudo-redshifts derived from the probability density distribution.

\section{Lynden-Bell's $c^{-}$ and Efron-Petrosian methods}\label{Lynden-Bell's $C^{-}$ method}
Observational truncation is inevitable due to the limited sensitivity of instruments. The Lynden-Bell's $c^{-}$ and Efron-Petrosian methods are effective for reconstructing the joint luminosity and redshfit distribution of objects from truncated data samples. The method can break the degeneracy between the luminosity function and the formation rate, and has been widely applied to quasars \citep{1971MNRAS.155...95L,1992ApJ...399..345E,petrosian1993interpretation,1999ApJ...518...32M}, GRBs \citep{2002ApJ...574..554L,2004ApJ...609..935Y,2012MNRAS.423.2627W,2015ApJS..218...13Y,2025arXiv251113783L}, and FRBs \citep{2019JHEAp..23....1D,2024ApJ...973L..54C,2025ApJ...988L..64C,2025ApJ...995L..53G}.

In this non-parametric method, it is assumed that the luminosity $L$ and redshift $z$ are independent. So that the joint distribution can be expressed as $\Psi(L, z) = \psi(L)\rho(z)$, where $\psi(L)$ represents the luminosity function and $\rho(z)$ means the comoving density rate of FRBs \citep{1992ApJ...399..345E}. However, the luminosity and redshift of FRBs are not independent \citep{2024ApJ...973L..54C,2025ApJ...988L..64C}. Here, we adopt an assumption of luminosity evolution $g(z)$, so that the transformed luminosity $L_0 = L/g(z)$ is independent of redshift. Therefore, we can use the non-parametric method to obtain $\Psi(L_0, z) = \psi(L_0)\rho(z)$. 

\begin{figure}
    \centering
    \includegraphics[width=0.7\textwidth]{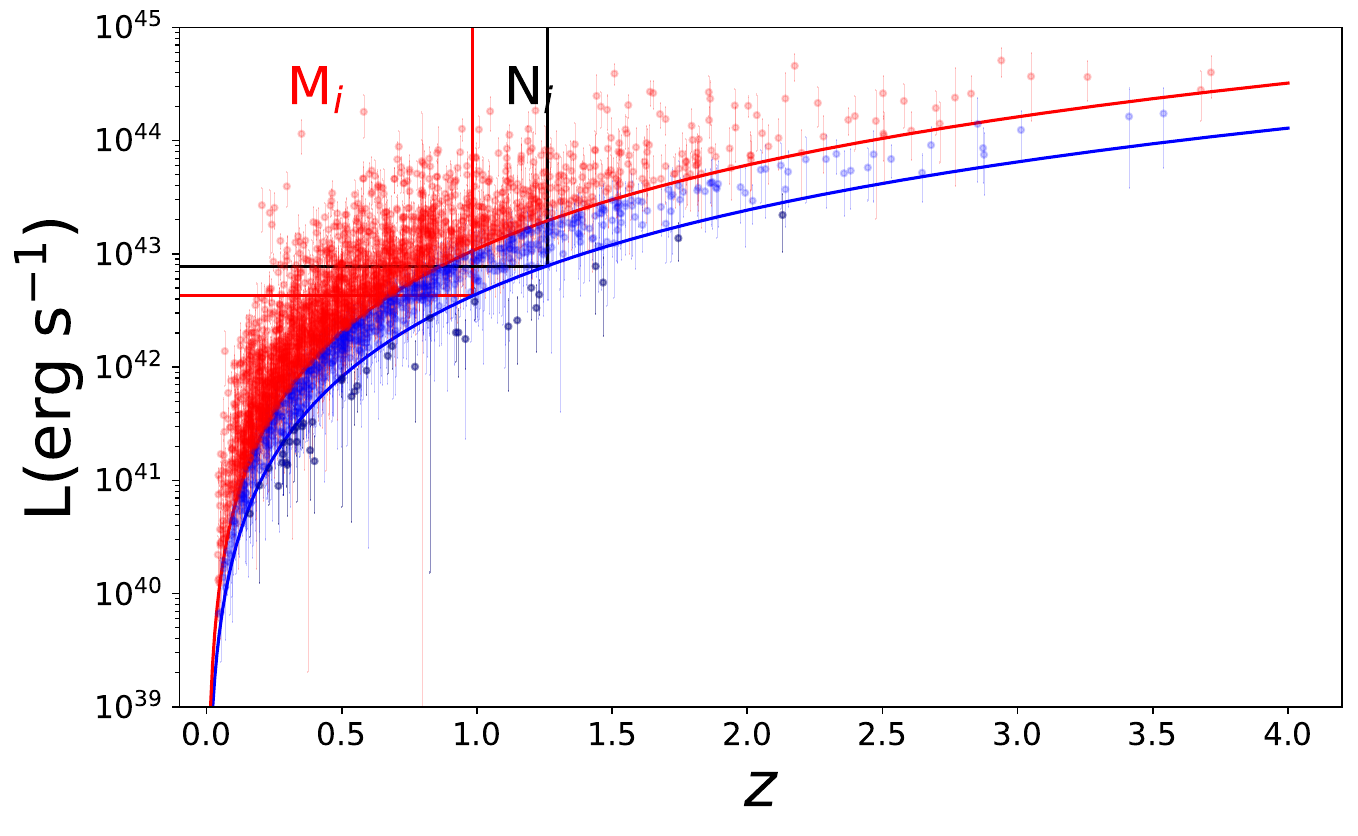} 
    \caption{Luminosity-redshift distribution of FRBs. We adopt the best-fit pseudo-redshift values as the redshifts of the FRBs. The blue solid line is the flux limit of $0.2$ Jy, while the red solid line corresponds to the flux limit of $0.5$ Jy. Black dots represent FRBs below the flux limit of $0.2$ Jy, blue dots represent FRBs between flux limits of $0.2$ Jy and $0.5$ Jy, and red dots represent FRBs above the flux limit of $0.5$ Jy. The error bars are derived by flux errors. $M_i$ and $N_i$ are shown in the red and black rectangle, respectively.}
    \label{Fig:z-L}
\end{figure}

Following the calculations in Section \ref{Calculating Redshift}, we obtain the redshifts and their $1\sigma$ uncertainties for the FRB sample. We can now use these results to compute the luminosities. Taking into account the uncertainties in the redshift distribution -- specifically, the best-fit value and its $1\sigma$ uncertainty -- we consider three distinct samples: (1) lower, all FRBs are assigned redshifts at the $1\sigma$ lower limit; (2) mean, all FRBs are assigned their best-fit redshift values; (3) upper, all FRBs are assigned redshifts at the $1\sigma$ upper limit. The luminosity can be expressed as
\begin{equation}
    L = 4\pi d_L^2(z) F \Delta\nu,
\end{equation}
where the luminosity distance is
\begin{equation}
    d_L(z) = \frac{c(1+z)}{H_0} \int_0^z \frac{dz}{\sqrt{1 - \Omega_m + \Omega_m(1+z)^3}}.
\end{equation}
The parameter $F$ denotes the peak flux detected by CHIME in a specific frequency interval spanning a bandwidth of $\Delta \nu = 400$MHz. Current observations of FRB spectra by CHIME are band-limited. The spectra are poorly constrained \citep{2026arXiv260109399F}.  Therefore, the $k$-correction for FRBs is not considered. For the flux limits, we select $0.2$ Jy and $0.5$ Jy, respectively. The calculated luminosities and redshifts of the sample are shown in Figure \ref{Fig:z-L}. We adopt the best-fit pseudo-redshift values obtained in Section \ref{Calculating Redshift} as the redshifts of the FRBs, and the luminosity uncertainties are propagated from the flux measurement errors. The blue and red curves in Figure \ref{Fig:z-L} represent the luminosities corresponding to flux limits of $0.2$ Jy and $0.5$ Jy, respectively. Black dots denote FRBs below the limit of $0.2$ Jy, blue dots represent FRBs with fluxes between $0.2$ Jy and $0.5$ Jy, and red dots indicate FRBs above the $0.5$ Jy flux limit.

In simple terms, the Lynden-Bell's $c^{-}$ method is a non-parametric approach for handling truncated data. It constructs an associated set for each observed source and assigns a weight based on the source's rank in luminosity or redshift within that sed, enabling an unbiased reconstruction of the luminosity function and redshift distribution. 
Here, we note that while the $g(z) = (1+z)^k$ law serves as a good approximation as a simple assumption \citep{2024ApJ...973L..54C}, it introduces significant deviations at high redshifts. Therefore, following the work of \cite{2025ApJ...988L..64C}, we adopt 
\begin{equation}
    g(z) = (1+z)^k \frac{1 + (1+z_{cr})^k}{(1+z)^k + (1+z_{cr})^k}.
\end{equation}
This function scales as $(1+z)^k$ at low redshifts but flattens out at high redshifts. Following \cite{2025ApJ...988L..64C}, we fix $z_{cr}$ as 2.5. For each point $(L_{i},z_{i})$, we can define an associated set $J_i$ as \citep{1992ApJ...399..345E}
\begin{equation}
J_{i}=\left\{j\mid L_j\geqslant L_i,z_j\leqslant z_i^{\max}\right\},
\end{equation}
where $L_i$ is the $i$th FRB luminosity and $z_i^{\max}$ is the maximum redshift where the FRB with luminosity $L_i$ can be detected. The region is shown in Figure \ref{Fig:z-L} as a black rectangle. The number of FRBs contained in this region is $n_i$. The number $N_i=n_i-1$ means taking the $i$th FRB out. 

Similarly, we can define $J_i^{\prime}$  as
\begin{equation}
J_{i}^{\prime}=\left\{j\mid L_{j}\geqslant L_{i}^{\mathrm{lim}},\:z_{j}<z_{i}\right\}.
\end{equation}
In this equation, $z_i$ denotes the redshift of the $i$th FRB and $L_i^\mathrm{lim}$ represents the minimum observable luminosity at that redshift. The number of events contained within this region is $M_i$. This region is displayed as the red rectangle in Figure \ref{Fig:z-L}.

\begin{figure}
    \centering
    \includegraphics[width=0.8\textwidth]{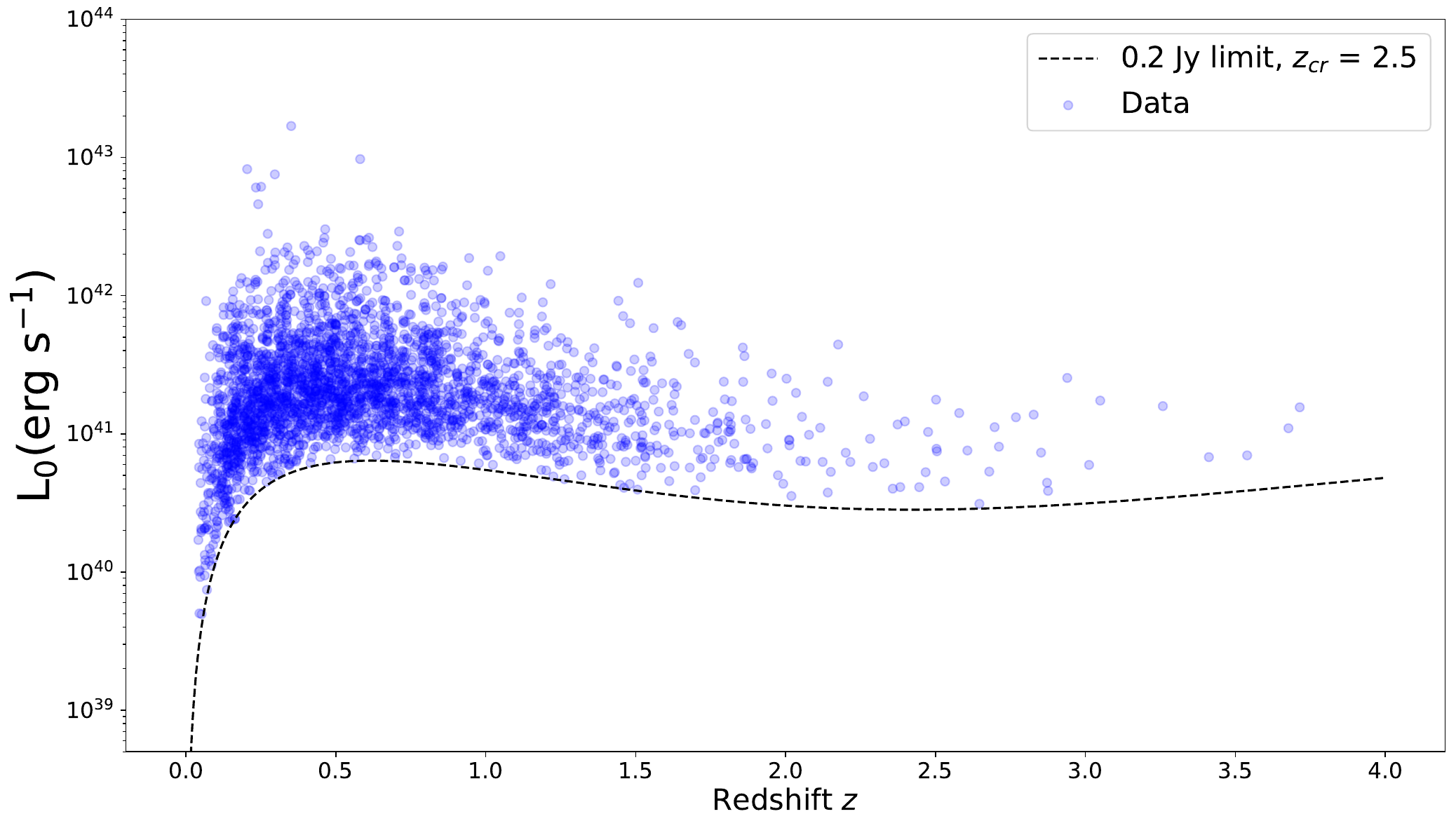} 
    \caption{Non-evolving luminosity $L_0 = L/g(z)$ under $z_{cr} = 2.5$. The blue dots represent FRBs in the mean sample. The dotted line means the flux limit.}
    \label{Fig:z-L0}
\end{figure}

\begin{figure}
    \centering
    \includegraphics[width=0.8\textwidth]{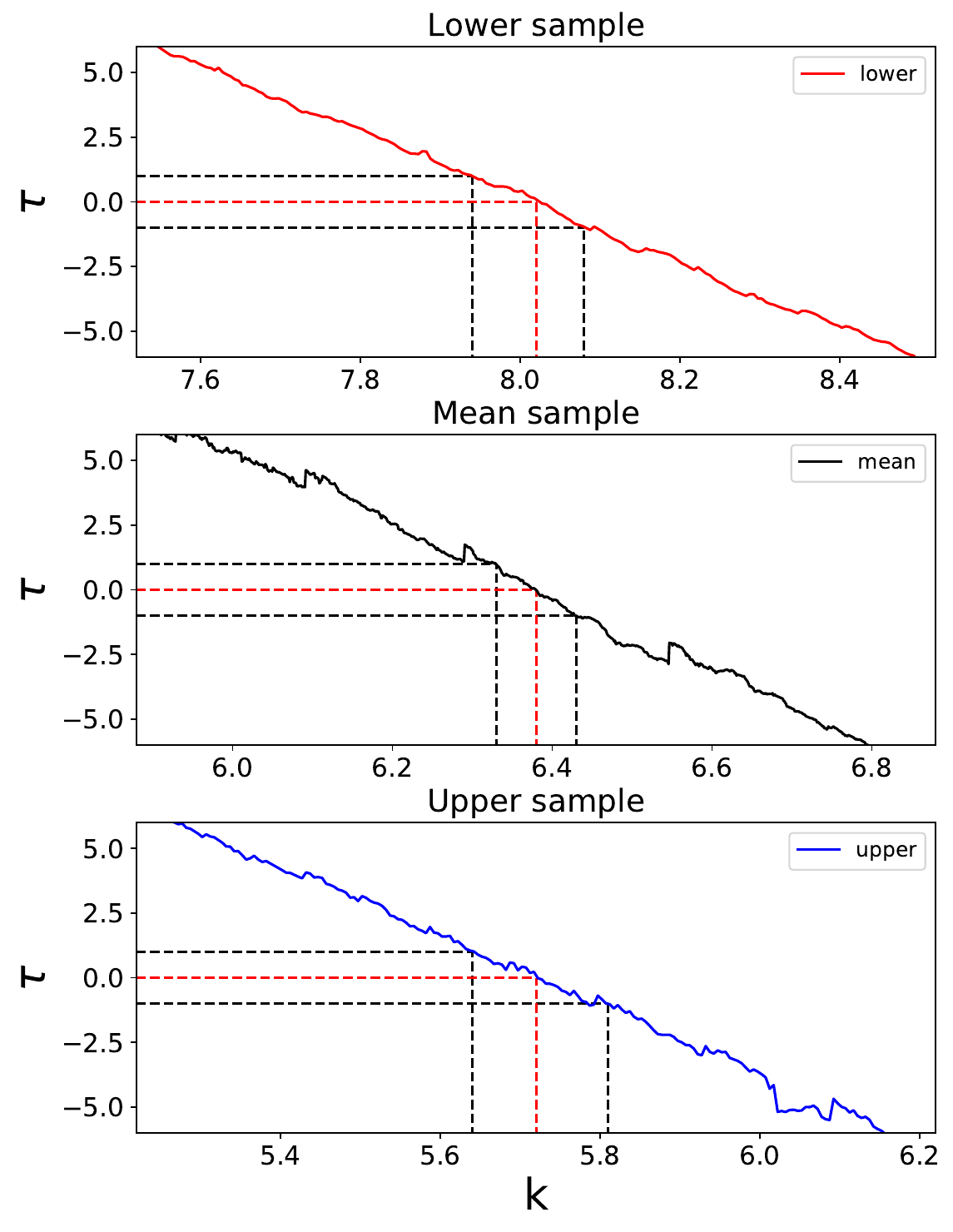} 
    \caption{Value of test statistic $\tau$ as a function of $k$. The red dotted line represents the best fit for $\tau = 0$ and the black dotted lines are the $1\sigma$ errors. The color of Red, black, blue denote the lower, mean, and upper redshift samples, respectively, with corresponding $k$ values of $8.02^{+0.06}_{-0.08}$, $6.38^{+0.05}_{-0.05}$, and $5.72^{+0.09}_{-0.08}$, along with their $1\sigma$ confidence level.}
    \label{Fig:tau}
\end{figure}

\begin{figure}
    \centering
    \includegraphics[width=0.9\textwidth]{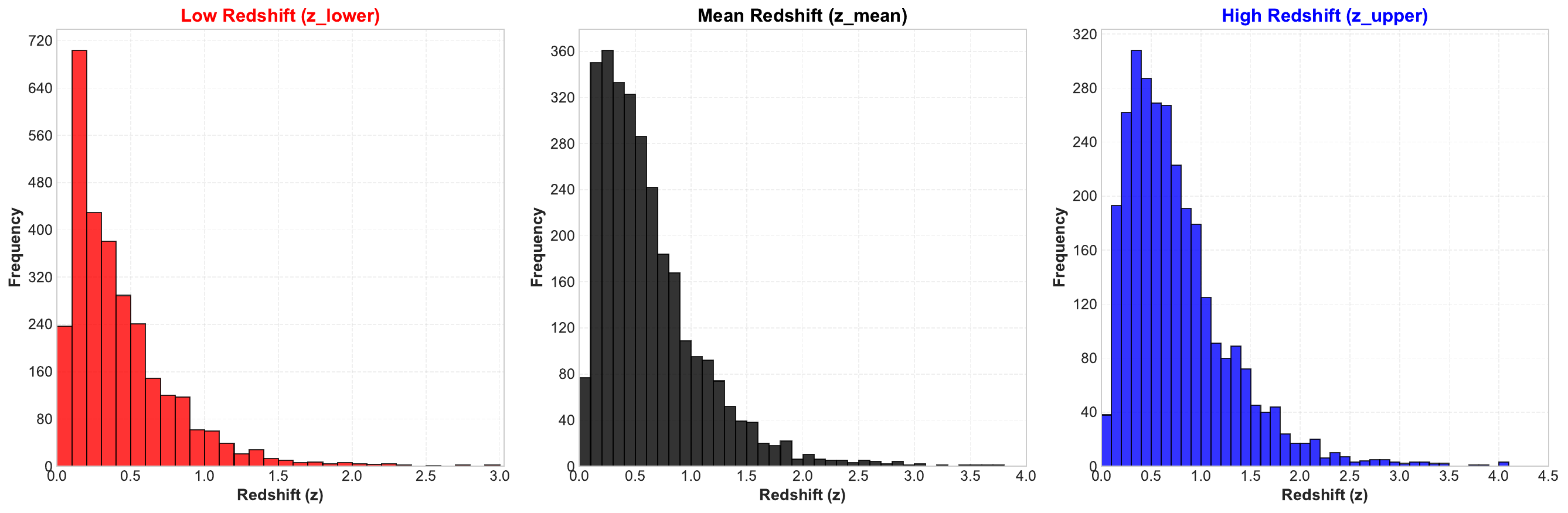} 
    \vspace{0.5cm} 
    \includegraphics[width=0.9\textwidth]{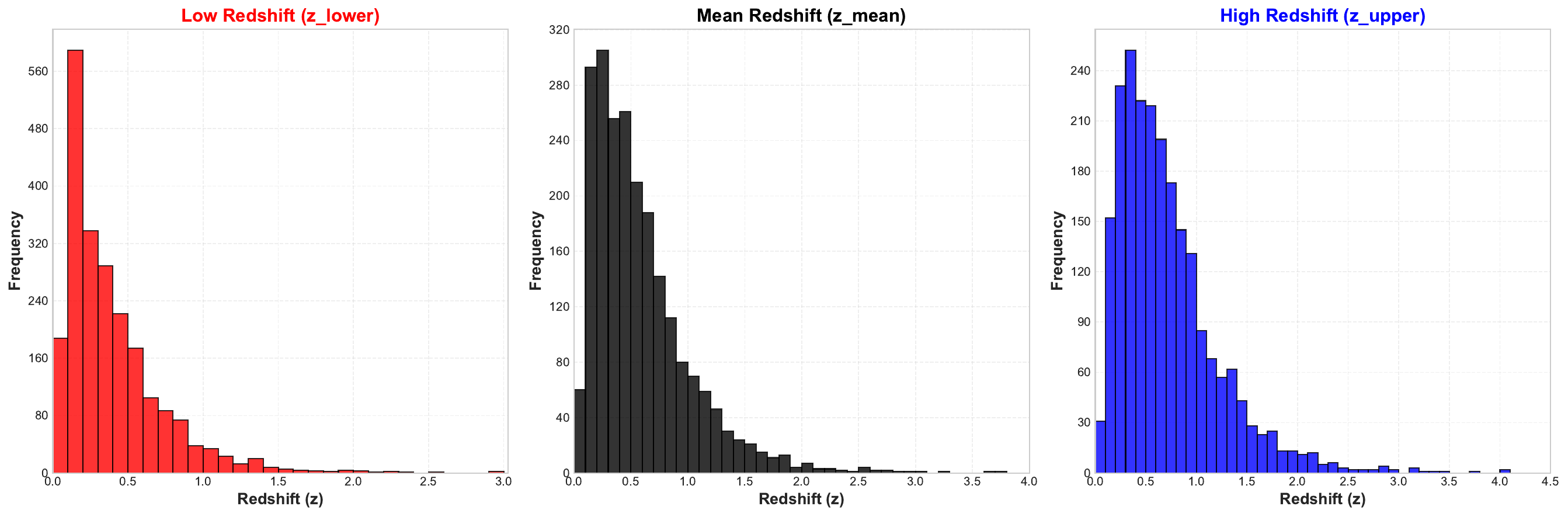}
    \caption{Redshift distribution of the FRB sample. The upper panel shows the redshift distribution of the FRB sample for the flux limit of $0.2$ Jy, corresponding to the lower bound, mean value, and upper bound of the redshift, respectively. The low panel displays the same for a flux limit of $0.5$ Jy.}
    \label{Fig:redshift_distribution}
\end{figure}

First, we analyze the $n_i$ FRBs located within the black rectangle in Figure \ref{Fig:z-L}. We define $R_i$ as the number of events with redshift less than or equal to $z_i$. If $L$ and $z$ are statistically independent, then $R_i$ should follow a uniform distribution between $1$ and $n_i$. Here, we define the test statistic $\tau$ as 
\begin{equation}
\tau=\sum_i\frac{(R_i-E_i)}{\sqrt{V_i}},
\end{equation}
where $E_i = \frac{1 + n_i}{2}$ and $V_i = \frac{n_i^2 - 1}{12}$ are the expected mean and the variance of $R_i$. When the test statistic $\tau$ approaches zero, $R_i$ is uniformly distributed between $1$ and $n_i$, indicating that luminosity and redshift are independent. Therefore, we can adjust the value of $k$ such that $\tau = 0$, thereby determining the form of $g(z)$. 
After obtaining the value of $k$, we compute the non-evolving luminosity $L_0$ using the relation $L_0 = L / g(z)$. We present the evolution of $L_0$ with redshift for different values of flux limits in Figure \ref{Fig:z-L0}. 
The result of the test statistic $\tau$ with $k$ under $z_{cr} = 2.5$ is shown in Figure \ref{Fig:tau}. For the sample of mean, the best-fit value of $k$ with $1\sigma$ confidence level is $6.38^{+0.05}_{-0.05}$. For the lower and upper samples, the values of $k$ are $8.02^{+0.06}_{-0.08}$ and $5.72^{+0.09}_{-0.08}$, respectively. Here we note that, when adopting the upper and mean redshift values, our results are consistent with those of \cite{2025ApJ...988L..64C} within $1 \sigma$ level. However, for the lower redshift bound, we obtain a significantly larger value of $k$. In Figure \ref{Fig:redshift_distribution}, we plot the redshift distributions of the samples under different redshift assumptions and flux limits. It is evident that, compared to the mean and upper samples, the lower sample exhibits a noticeably different distribution, with a clear concentration at low redshifts. This concentration likely explains why a larger $k$ value is obtained for the lower sample. Here we only present the three different samples for the case with a flux limit of $0.2$ Jy. The results for the flux limit of $0.5$ Jy are shown in Table \ref{T_BPL}. Here we note that different choices of the redshift $z$ lead to different redshift evolution indices $k$. The results for the lower and upper cases are not symmetrically distributed around the mean sample, which may be due to the fact that the lower and upper bounds of the pseudo-redshift derived for the FRBs are asymmetric. As the redshift value increases from lower to mean and then to upper, the value of $k$ gradually decreases. When a stricter flux limit $0.5$ Jy is applied, the value of $k$ remains nearly unchanged. 

\section{Luminosity function and Formation rate}\label{Luminosity function and Formation rate}
After determining the value of $k$, we can calculate the cumulative luminosity function and the cumulative redshift distribution. From the non-parametric method, the cumulative luminosity function $\psi(L_0)$ can be expressed as \citep{1971MNRAS.155...95L,1992ApJ...399..345E}
\begin{equation}
\psi(L_{0i})=\prod_{j<i}(1+\frac1{N_j}),
\end{equation}
where $j<i$ means that the FRB has a luminosity $L_{0j}$ larger than $L_{0i}$. The cumulative redshift distribution $\phi(z)$ is
\begin{equation}
\phi(z_i)=\prod_{j<i}(1+\frac1{M_j}),
\end{equation}
where $j<i$ means that the FRB has redshift $z_j$ less than $z_i$. 

\begin{figure}
    \centering
    \includegraphics[width=0.6\textwidth]{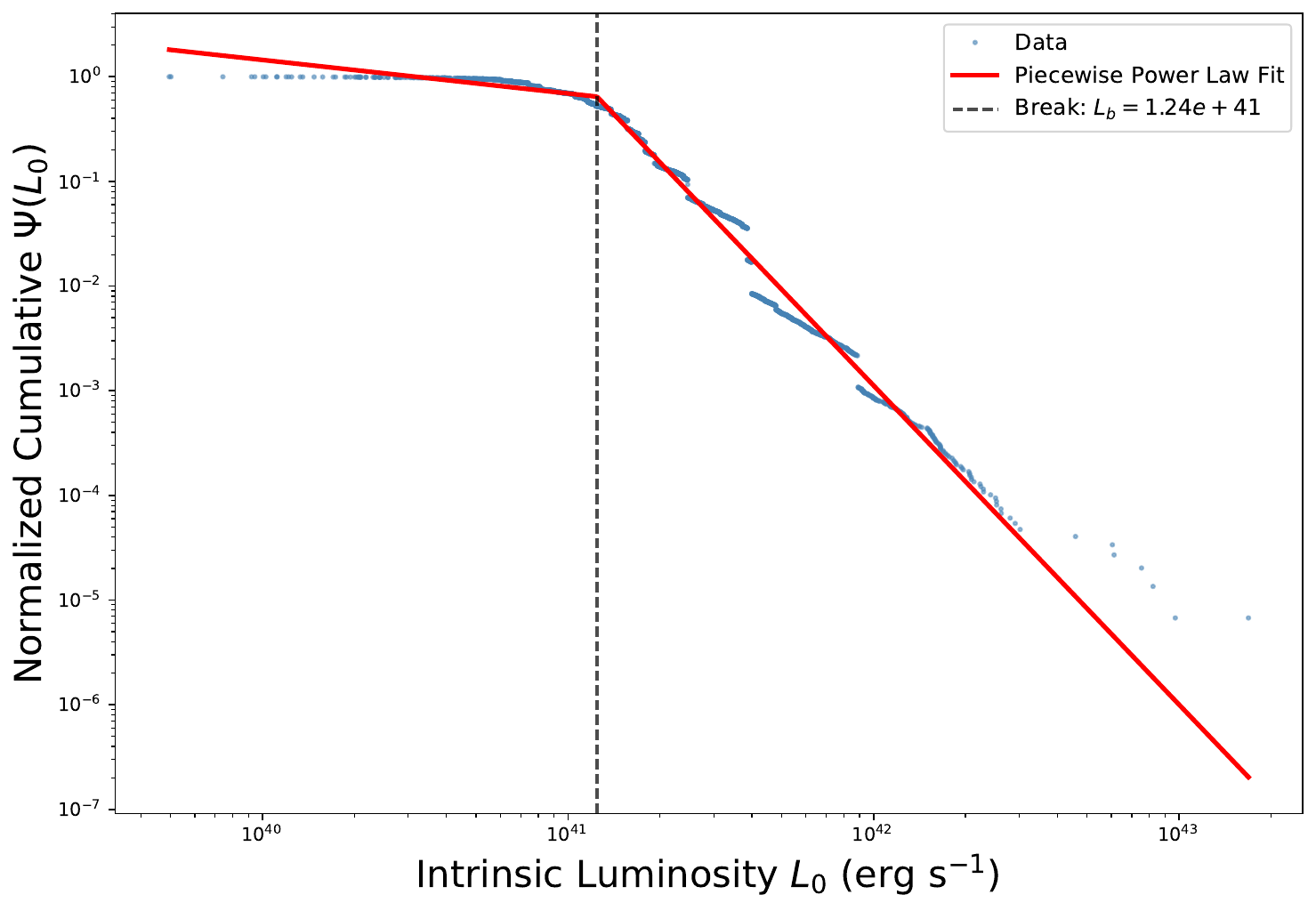} 
    \caption{Cumulative luminosity function $\psi(L_0)$, which is normalized to unity at the lowest luminosity. The red line is the best fit with a broken power-law model. The result is obtained using the mean redshift under a flux limit of $0.2$ Jy.} \label{Fig:Cumulative_Luminosity_with_fit}
\end{figure}

The results of cumulative luminosity function are shown in Figure \ref{Fig:Cumulative_Luminosity_with_fit}. Here, we do not show results for all combinations of sample and flux limits; instead, we present only the case with a flux limit of $0.2$ Jy for the mean sample as an illustration. The luminosity function $\psi(L_0)$ can be fitted with a broken power law model. The best fit for dim and bright bursts is given by
\begin{equation}
\psi(L_0)\propto\begin{cases}L_0^{-0.32\pm0.02}&L_0<L_0^b\\L_0^{-3.05\pm0.01}&L_0>L_0^b\end{cases},
\label{luminosity}
\end{equation}
where $L_0^b=  1.24 \times 10^{41}$ erg s$^{-1}$ is the break luminosity point. The detailed results are presented in Table \ref{T_BPL}. It is worth noting that there are no significant differences among the results for the different samples. As can be seen in the case with a flux limit of $0.2$ Jy, the luminosity function parameters $\alpha$ and $\beta$ remain nearly identical across different choices of redshift. When the flux limit is set to $0.5$ Jy, the values of the parameters $\alpha$ and $\beta$ remain consistent across different redshift samples. However, compared to the case with a flux limit of $0.2$ Jy, some fainter FRBs are excluded due to the stricter cut. Consequently, the dim parameter $\alpha$ at a flux limit of $0.5$ Jy is smaller than that at a flux limit of $0.2$ Jy. Although the values of the break luminosity point $L^b_0$ differ slightly among the samples, this is expected because the calculation of $L$ explicitly depends on $z$. Consequently, the choice of redshift shifts the entire sample systematically, leading to different $L^b_0$ values.

As shown in Figure \ref{Fig:Cumulative_Luminosity_with_fit}, the break luminosity for millisecond-duration radio bursts is $1.24 \times 10^{41}$ erg s$^{-1}$, corresponding to a break energy of $\sim 10^{38}$\,erg for FRBs in the CHIME/FRB Catalog 2. \cite{Wu2025ApJ} investigated the energy distributions of three hyper-active repeating FRBs observed by the Five-hundred-meter Aperture Spherical radio Telescope (FAST) and reported a universal break energy at approximately $10^{38}$\,erg. The results we obtained in this work using non-repeating FRBs show remarkable consistency with those derived by \cite{Wu2025ApJ} using repeating FRBs. Such consistency suggests that both inactive and active FRBs may share a common physical origin, potentially involving a starquake-triggering mechanism in neutron stars. Our results further support the conclusion of \cite{Beniamini2025} and \cite{2025ApJ...989..144L} -- based on the relationship between volumetric rate and burst energy -- that repeating and non-repeating FRBs share a common physical origin.

\begin{table*}
	\caption{The results obtained by using different redshift samples under various flux limit scenarios are listed in this table. The parameter $k$ is the index of $1+z$ in the function of $g(z)$. The parameter $L^b_0$ is the break luminosity point. Parameters $\alpha$ and $\beta$ are the faint-end and bright-end power-law indices, respectively, in Equation \ref{luminosity}. The last column lists the parameter $\gamma$, which is the index of $(1+z)$ in Equation \ref{eq:rou}. \label{T_BPL}}
	\centering
	\begin{tabular}{ccccccc}
		\\
		\hline
		Flux Limit	&  $z$  &  $k$  & $L_0^b$      & $\alpha$              & $\beta$  &  $\gamma$                \\
		\hline
             &lower & $8.02^{+0.06}_{-0.08}$   & $8.3\times10^{40}$  &  $0.27 \pm 0.01$ & $1.59 \pm 0.01$    &  $5.68 \pm 0.06$\\
        0.2 Jy  &mean  & $6.38^{+0.05}_{-0.05}$   & $1.24\times10^{41}$ &  $0.32 \pm 0.02$ & $3.05 \pm 0.01$ & $5.38 \pm 0.02$\\
             &upper & $5.72^{+0.09}_{-0.08}$   & $1.56\times10^{41}$ &  $0.19 \pm 0.02$ & $2.82 \pm 0.01$    & $4.74 \pm 0.02$\\ \hline
             &lower & $8.36^{+0.08}_{-0.11}$   & $8.72\times10^{40}$  &  $0.26 \pm 0.07$ & $1.75 \pm 0.01$     & $6.28 \pm 0.06$\\
        0.5 Jy  &mean  & $6.55^{+0.04}_{-0.14}$   & $1.16\times10^{41}$ &  $0.20 \pm 0.04$ & $3.66 \pm 0.01$ & $5.64 \pm 0.03$\\
             &upper & $5.31^{+0.42}_{-0.61}$   & $2.45\times10^{41}$ &  $0.48 \pm 0.06$ & $5.83 \pm 0.02$    & $4.54 \pm 0.03$\\
                      
		\hline
	\end{tabular}
\end{table*}

\begin{figure}
    \centering
    \includegraphics[width=0.6\textwidth]{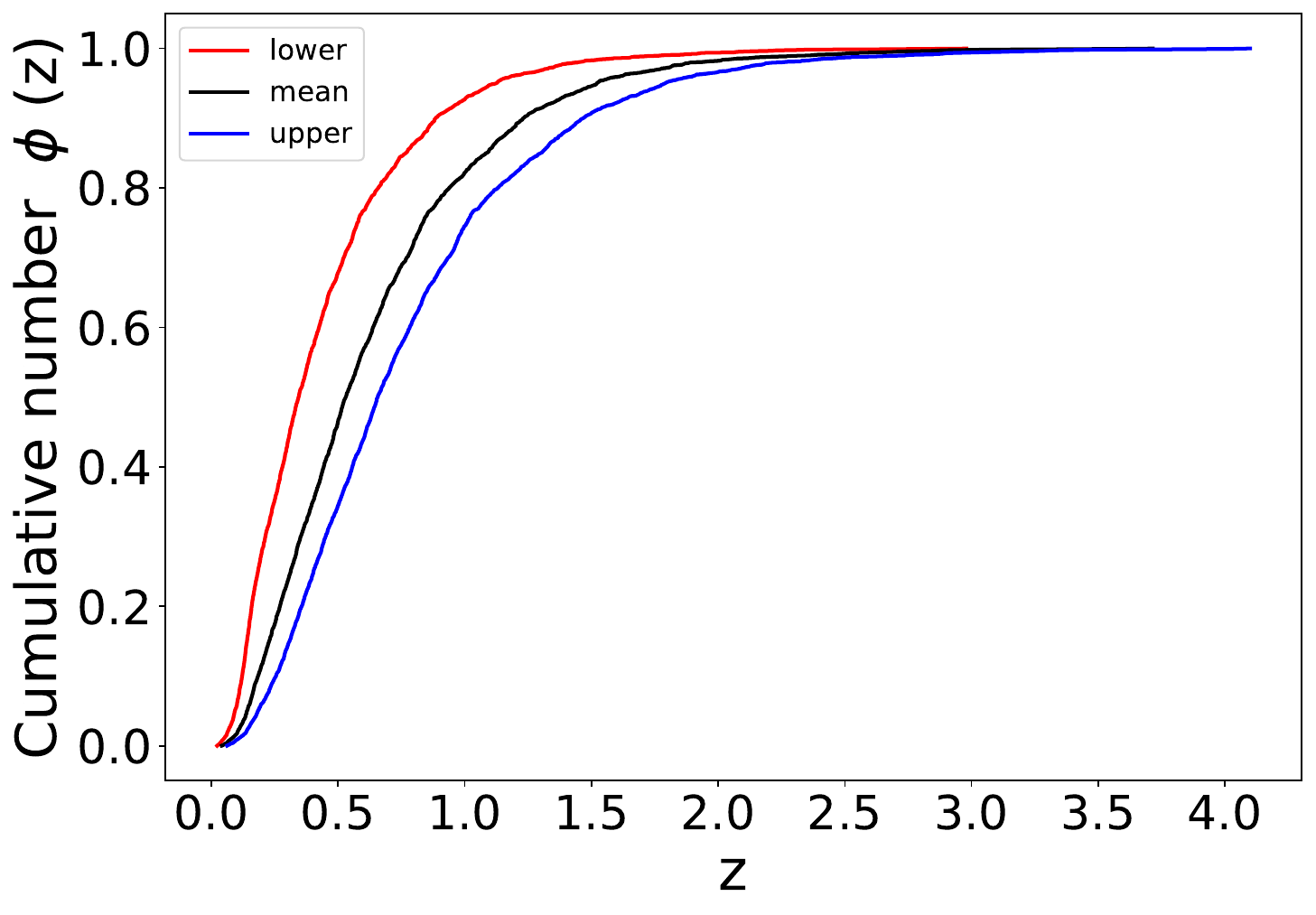} 
    \caption{Normalized cumulative redshift distribution of FRBs. The color of red, black, and blue represent the lower, mean, upper redshift samples, respectively.}
    \label{Fig:Cumulative_z}
\end{figure}

\begin{figure}
    \centering
    \includegraphics[width=0.8\textwidth]{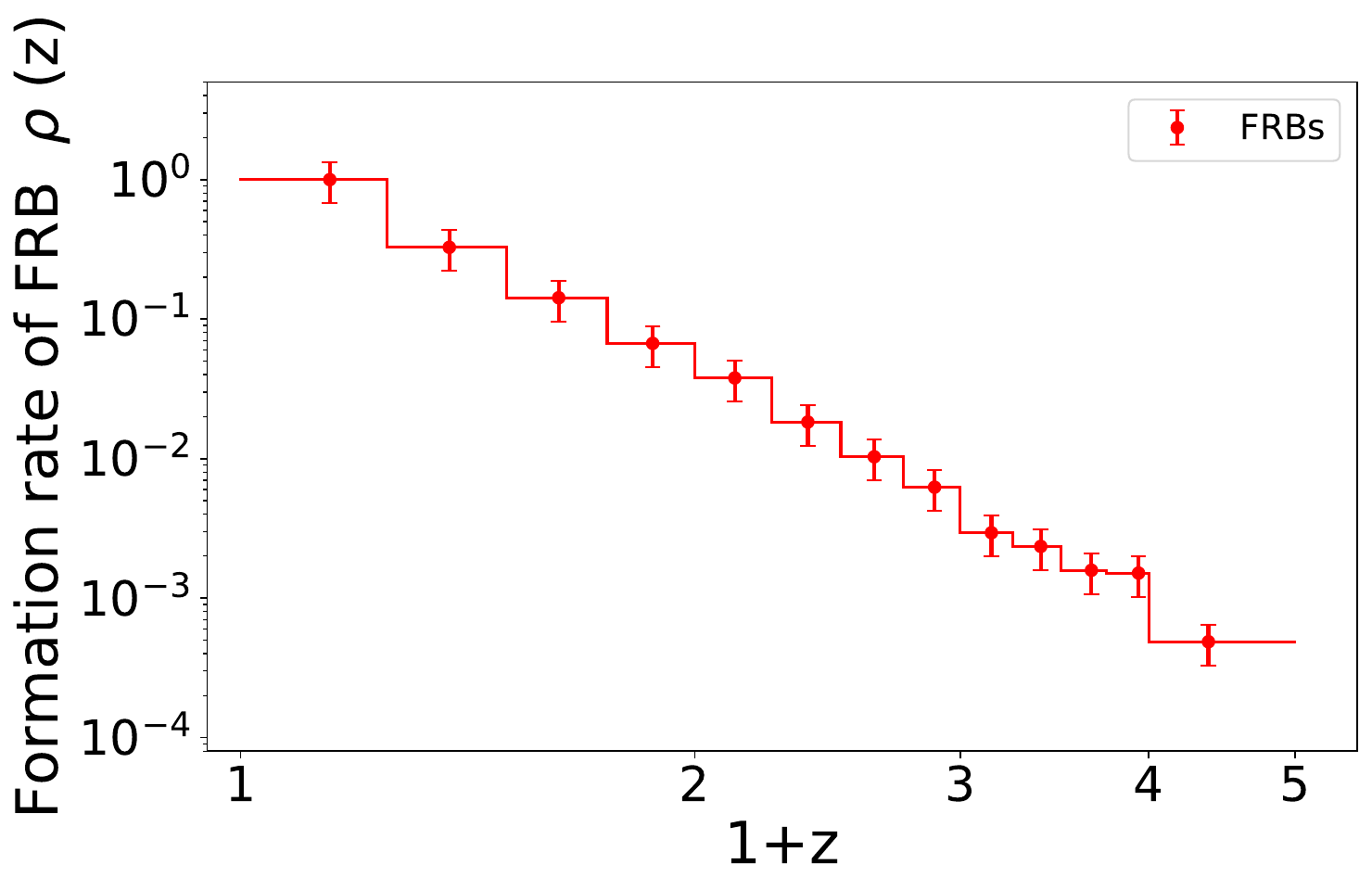} 
    \caption{Comoving formation rate of FRBs, which is normalized to unity at the first point. The $1\sigma$ error is also shown.}
    \label{Fig:Formation_rate_FRB}
\end{figure}

The cumulative redshift distribution $\phi(z)$ are shown in Figure \ref{Fig:Cumulative_z}. We present the results for the case with a flux limit of $0.2$ Jy, considering three redshift scenarios: lower, mean, and upper. Based on this, we can further derive the formation rate of FRBs as
\begin{equation}
\rho(z)=\frac{d\phi(z)}{dz}(1+z)\Bigg(\frac{dV(z)}{dz}\Bigg)^{-1}.
\end{equation}
The factor $(1+z)$ results from the cosmological time dilation and the differential comoving volume $dV(z)/dz$ can be expressed as
\begin{equation}
\begin{aligned}
\frac{dV(z)}{dz}&=4\pi\Bigg(\frac{c}{H_0}\Bigg)^3\Bigg(\int_0^z\frac{dz}{\sqrt{1-\Omega_{\mathrm{m}}+\Omega_{\mathrm{m}}(1+z)^3}}\Bigg)^2  \\
&\times\frac1{\sqrt{1-\Omega_{\mathrm{m}}+\Omega_{\mathrm{m}}(1+z)^{3}}}.
\end{aligned}
\end{equation}
The results are shown in Figure \ref{Fig:Formation_rate_FRB}. Here, the redshifts used are the mean values. We adopt a redshift bin width of $\Delta z = 0.25$, but for $z >3$, the number of data points become small; therefore, the last bin is widened to a width of $1$. The formation rate can be fitted by a power-law function as 
\begin{equation}
\rho(z) \propto (1 + z)^{-5.38 \pm 0.02}. \label{eq:rou}
\end{equation}
The result is consistent with the result of \cite{2024ApJ...973L..54C} and \cite{2025ApJ...988L..64C}, but yields a steeper power-law index. The results corresponding to different flux limits and different redshift selections are presented in Table \ref{T_BPL}, with the parameter $\gamma$ given in the last column. As can be seen from the results, as the redshift $z$ selection changes from lower to mean to upper, the index $\gamma$ gradually decreases, indicating that the decline rate becomes increasingly shallower. When the flux limit is set to $0.5$ Jy, the index $\gamma$ is consistently larger than when the flux limit is set to $0.2$ Jy. The rapid decline of FRB rate is robust in all cases.

\begin{figure}
    \centering
    \includegraphics[width=0.8\textwidth]{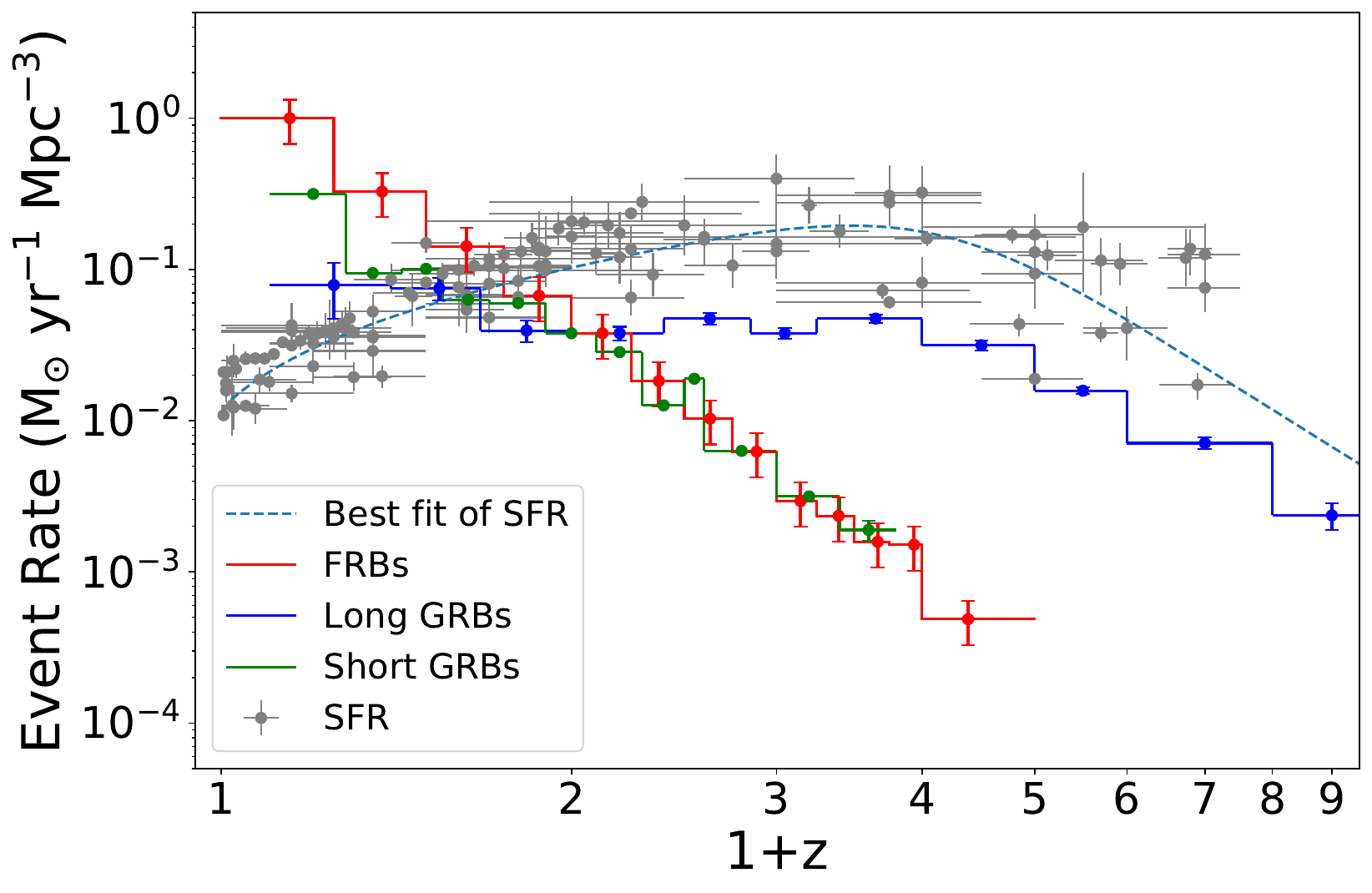} 
    \caption{Comparison of the FRB formation rate (red) with other events. The green and blue lines represent the rates of short GRBs \citep{2018ApJ...852....1Z} and long GRBs \citep{2015ApJS..218...13Y}, respectively. Gray dots and blue dashed line correspond the observed SFR and the best fit \citep{2006ApJ...651..142H} 
    }
    \label{Fig:SFR}
\end{figure}

We compare the FRB formation rate with SFR \citep{2006ApJ...651..142H}, which can help shed light on the origin of FRBs. If FRBs originate from the death of massive stars, their formation rate should closely trace the SFR. The results are shown in Figure \ref{Fig:SFR}. It can be seen that there is a clear time delay between the FRB rate and the SFR. We also compare our result with long gamma-ray bursts (GRBs) \citep{2015ApJS..218...13Y} and short GRBs \citep{2018ApJ...852....1Z}. The redshift-dependence of FRB rate is similar as short GRBs. Our comparison suggests that non-repeating FRBs may be primarily associated with old populations rather than young populations traced by SFR. Our results indicate that old neutron stars or black holes are more likely to be their progenitors.

\section{Conclusions and discussion}\label{Conclusions and discussion}
In this work, we analyze the recently released CHIME catalog 2 data. After applying data selection criteria, we compute pseudo-redshift for the FRBs. We then employ the Lynden-Bell’s $C^{-}$ and Efron-Petrosian methods to break the degeneracy between the cumulative luminosity function and the cumulative redshift distribution of FRBs. By removing the redshift evolution, we derive the luminosity function and redshift distribution. Finally, we compare the FRB event rate with the SFR and find a significant time delay.
Our main conclusions can be summarized as follows:
\begin{itemize}
\item We adopt different flux limit and redshift assignment schemes. Compared to the case with a flux limit of $0.2$ Jy, the value of $k$ is larger under the flux limit of $0.5$ Jy. Regarding the redshift choice, as $z$ increases from the lower to mean to upper, the corresponding $k$ gradually decreases. We note that the results for the lower and upper cases are not symmetrically distributed around the mean sample -- this asymmetry arises because the lower and upper bounds of the pseudo-redshift derived for the FRBs are themselves asymmetric. 

\item For the luminosity function after removing redshift evolution, we find that it is well described by a broken power-law model. Across different flux limits and redshift samples, the power-law indices of the luminosity function show remarkable consistency. This consistency also supports that our results is robust, independent of choice of redshfits and flux limits. It is worth noting that the derived values of the break luminosity point $L_0^b$ differ among the different redshift samples. This is because the calculation of the luminosity $L$ explicitly depends on the redshift $z$, which systematically affects the resulting $L$ values and consequently leads to different best-fit $L_0^b$ values in the subsequent analysis.

\item We find that the break luminosity point $L_0^b$ of the FRB luminosity function is approximately $10^{41}$ erg s$^{-1}$, corresponding to a break energy of $\sim 10^{38}$\,erg for milli-second bursts. This result is consistent with the findings of \cite{Wu2025ApJ} based on three hyper-active repeating FRBs observed by FAST. Such consistency suggests that both inactive and active FRBs may share a common physical origin. 

\item The formation rate of FRBs can be fitted by a power-law function as $\rho(z) \propto (1 + z)^{-5.38 \pm 0.02}$. By comparing the FRB formation rate with SFR, we find a clear time delay between them. Our comparison suggests that non-repeating FRBs may be primarily associated with old populations rather than young populations traced by SFR. Our results indicate that magnetars are the progenitors of FRBs.

\end{itemize}

\begin{acknowledgments}
We thank the referee for valuable comments and suggestions, which have helped to improve this manuscript. This work was supported by the National Natural Science Foundation of China (grant Nos. 12494575 and 12273009). Qin Wu is supported by National Natural Science Foundation of China (NSFC) (grant Nos. 12447115 and 12503050) and the China Postdoctoral Science Foundation (CPSF) under Grant Number GZB20240308, 2025T180875 and 2025M773199. 

\end{acknowledgments}

\dataset[]{https://zenodo.org/records/18843430}

\bibliography{sample701}{}
\bibliographystyle{aasjournalv7}

\clearpage

\end{document}